# Fundamental hydrogen storage properties of TiFe-alloy with partial substitution of Fe by Ti and Mn

Erika Michela Dematteis,[a,d)*] David Michael Dreistadt,[b)] Giovanni Capurso,[b)] Julian Jepsen,[b,c)] Fermin Cuevas,[a)] and Michel Latroche[a)]

[a)] Univ Paris Est Creteil, CNRS, ICMPE, UMR 7182, 2 rue Henri Dunant, 94320 Thiais, France

[b)] Institute of Materials Research, Helmholtz-Zentrum Geesthacht, Max-Plank-Str. 1, 21502 Geesthacht, Germany

[c)] Institute of Materials Technology, Helmut-Schmidt-University, Holstenhofweg 85, 22043 Hamburg, Germany

[d)] Present address: Department of Chemistry, Inter-departmental Center Nanostructured Interfaces and Surfaces (NIS), and INSTM, University of Turin, Via Pietro Giuria 7, 10125 Torino, Italy

*Corresponding author: Erika Michela Dematteis

E-mail address: erikamichela.dematteis@unito.it

**Graphical Abstract**

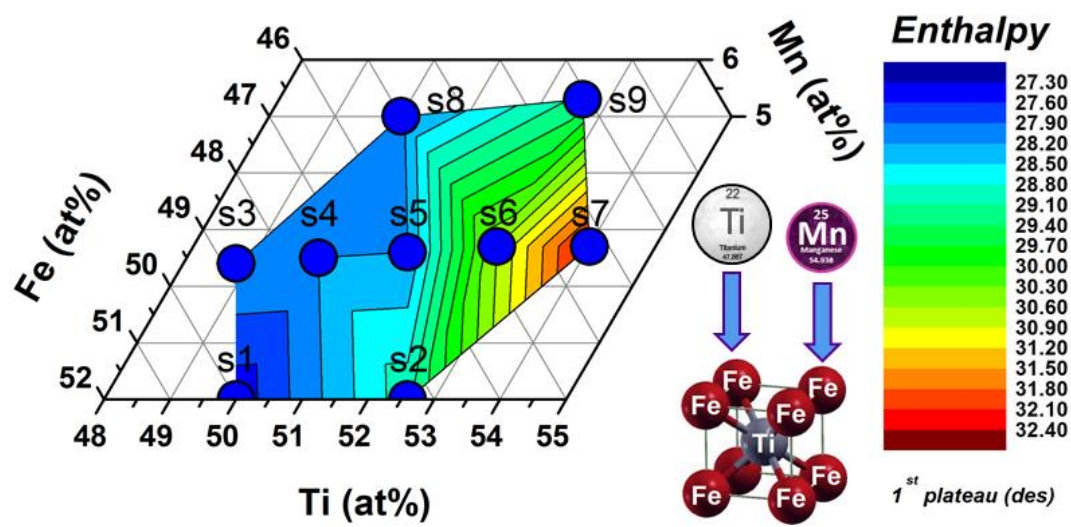

**Abstract**

TiFe intermetallic compound has been extensively studied, owing to its low cost, good volumetric hydrogen density, and easy tailoring of hydrogenation thermodynamics by elemental substitution. All these positive aspects make this material promising for large-scale applications of solid-state hydrogen storage. On the other hand, activation and kinetic issues should be amended and the role of elemental substitution should be further understood. This work investigates the thermodynamic changes induced by the variation of Ti content along the homogeneity range of the TiFe phase (Ti:Fe ratio from 1:1 to 1:0.9) and of the substitution of Mn for Fe between 0 and 5 at.%.

In all considered alloys, the major phase is TiFe-type together with minor amounts of $TiFe_2$ or $\beta$-Ti-type and $Ti_4Fe_2O$-type at the Ti-poor and rich side of the TiFe phase domain, respectively. Thermodynamic data agree with the available literature but offer here a comprehensive picture of hydrogenation properties over an extended Ti and Mn compositional range. Moreover, it is demonstrated that Ti-rich alloys display enhanced storage capacities, as long as a limited amount of $\beta$-Ti is formed. Both Mn and Ti substitutions increase the cell parameter by possibly substituting Fe, lowering the plateau pressures and decreasing the hysteresis of the isotherms. A full picture of the dependence of hydrogen storage properties as a function of the composition will be discussed, together with some observed correlations.

# 1. Introduction

Hydrogen as an energy carrier is increasingly coming into focus in politics and industry, as part of the promotion of renewable energies for environmental protection. Among other solutions, metal hydrides are suitable for storing hydrogen, by chemically storing and releasing it under appropriate thermodynamic conditions. For an industrial application, the technical efforts and costs required for the integration of such storage systems into energy supply systems are of critical importance. Room-temperature forming hydrides, such as TiFe-type alloys, which are able to absorb and desorb hydrogen at relatively low temperature (below 100 °C) and gas pressure (below 10 MPa), are suitable for this purpose.[1]

Starting from the 1970s, the use of low-temperature metal hydrides in stationary hydrogen storage[2,3] and the thermodynamic properties of TiFe alloys[4] have been investigated. Since then, extensive research has been carried out on the properties of TiFe compounds with various additional elements. Recently, Sujan *et al.*[5] have surveyed the properties and investigations of stoichiometric TiFe, while Dematteis *et al.* discussed in a comprehensive review the effects of elemental substitution on hydrogenation properties in TiFe-based alloys towards real applications.[6]

The development of TiFe alloys for industrial applications aims at achieving the highest possible storage capacity exploitable for stationary storage, but possibly promising for mobile application as well.[7] At the same time, alloy tailoring towards the most suitable equilibrium pressure is a key parameter for the technical application of a stationary energy storage system containing an electrolyzer and a fuel cell.[8] As a matter of fact, the low cost of TiFe-type alloys makes them promising for the integration into renewable energy systems and for grid balancing, as efficient energy storage media.[9]

Another crucial aspect in the use of TiFe compounds is the activation of the material. For pure equiatomic TiFe, it requires thermal annealing of the material up to 400 °C under hydrogen atmosphere.[10,11] These conditions require an increased technical effort, which should therefore be kept to a minimum. Instead of thermal treatments for activation, mechanical treatments like milling can be used.[12]

TiFe crystallizes in the CsCl-type cubic structure (*s.g. Pm-3m*). Upon hydrogenation, Pressure-Composition-Isotherms (PCI) are characterized by two subsequent plateau pressures related to the consecutive formation of the monohydride, β-TiFeH (orthorhombic, *s.g. P222$_1$*), and the dihydride, γ-TiFeH$_2$ (orthorhombic, *s.g. Cmmm*).[9,13–17] The occurrence of a double plateau implies larger pressure range for the usage of TiFe in practical application, and since the second plateau is found at too high pressure, it limits the exploitation of this material for many applications and it reduces significantly the reversible capacity in a limited pressure range.

TiFe exists in a small range of composition, which extends from 49.7 to 52.5 at.% Ti at 1000 °C.[5] Non-stoichiometry in TiFe-alloys are accommodated by partial substitution of Ti on Fe site.[18] Thus, the lattice parameter is expected to change within the homogeneity domain of TiFe due to the different atomic radius of Ti and Fe: $r_{Ti}$(1.46 Å) > $r_{Fe}$(1.24 Å).[19] In addition, activation and hydrogenation properties are affected as well. In fact, the Ti-rich compound TiFe$_{0.90}$ (*i.e.* Ti$^{1b}$(Ti$_{0.053}$Fe$_{0.947}$)$^{1a}$) requires almost no activation process for the first hydrogenation.[20]

The TiFe system was investigated in ref.[4], in the range between 33.3 and 66.7 at.% Ti (from 1:2 to 2:1 Ti:Fe ratio). At the iron rich side, TiFe$_2$ (hexagonal, *s.g. P63/mmc*) is present, which does not absorb hydrogen. The stoichiometric 1:1 compound TiFe has a capacity of 1.75 wt.% at room temperature. Ti$_{1+x}$Fe alloys with a higher titanium content can absorb hydrogen to a certain extent and Ti precipitates as secondary phase are present.[4] β-Ti precipitates (cubic, *s.g. Im-3m*) can be observed with a substantial dissolution of Fe, *i.e.* β-Ti$_{80}$Fe$_{20}$ phase.[10,21,22] With further increase of Ti content and in case of oxygen contamination either from starting materials or during synthesis and processing, compositions like Ti$_4$Fe$_2$O and Ti$_4$Fe$_2$O$_x$-type appear (cubic phase, *s.g. Fd-3m*), which can absorb hydrogen, *e.g.* to an extent equal to 1.58 wt.% in the case of Ti$_4$Fe$_2$O$_{0.25}$H$_{4.9}$.[23–25]

Beside the variation of the Ti:Fe ratio in pure TiFe, elemental substitutions can change significantly activation processes and hydrogen storage properties too.[14] Depending on alloy composition and based on Ti-Fe-*M* phase diagrams (where *M* is a substituting element), formation of secondary phase precipitates considerably influences the activation process and hydrogenation properties.[26] The microstructure of the alloy has a marked influence on hydrogenation and activation properties. Thus, if secondary phases precipitate at the grain boundaries, and especially if secondary phases react with hydrogen (as in the case of Ti to form $TiH_2$), activation and hydrogenation kinetics can be improved.[26]

Elemental substitution can possibly modify the stability of both plateau pressures in TiFe, as a function of the substituted amount.[27] The addition of substitutional elements can influence and reduce the gap between the two plateau pressures. However, in many cases the substitution can cause severe drawbacks, as in the case of aluminium (Al) that leads to a drastic reduction of the achievable hydrogen storage capacity.[28] Zirconium (Zr), chromium (Cr) and nickel (Ni) are other examples of elements used to influence the thermodynamic properties, but they can also cause a decrease in the gravimetric capacity.[5] Recently, it has been reported that the introduction of Cu in Mn-substituted alloy lowered the first plateau, while moving to higher pressure the second one, hence reducing sensibly the reversible capacity in a narrow pressure range.[26] Another important aspect in the choice of alloying elements is their availability and the dependence on metal supplies from volatile markets. For an industrial scale application of TiFe alloys, materials such as V or Y, should be possibly avoided.[29] The effect of other elements in the TiFe compound was investigated in the past to improve thermodynamic properties and hydrogen storage capacities. Co, for example, allows achieving a storage capacity as high as 1.98 wt.% at 0 °C and 10 MPa in the alloy $TiFe_{0.86}Mn_{0.05}Co_{0.05}$.[30]

Enlarging the TiFe cell parameter by elemental substitution might be useful for the hydrogenation properties of TiFe, since it generally results in lowering plateau pressure and improving activation in mild conditions, as in the case of Mn. The introduction of Mn, a non-critical raw material for Europe,[31] can also preserve good capacity and kinetics.[32] In the literature, many studies have been performed to describe in detail the thermodynamics and structure-substitution-properties relationships as a function of Mn content in Ti(Fe,Mn) compounds. Even before 1982, the substitution of Fe by Mn into TiFe was investigated to influence the thermodynamic properties and to reduce the activation issues of equiatomic TiFe.[18,24,33] At that time, it was already established that partial substitution of Fe by Mn does not change the crystal structure of TiFe, but expands it owing to the increased atomic size of Mn ($r_{Mn}$ = 1.35 Å ) as compared to Fe ($r_{Fe}$ = 1.24 Å).[19,24] The hydride phase is stabilized due to the increase of the binding energy.[33] A further important aspect is that the addition of Mn allows activation at a much lower temperature than with equiatomic TiFe,[5,10] thereby reducing the technical effort.

**Figure 1** shows the ternary Ti-Fe-Mn phase diagram at 1000 °C, where compositional domains in which the intermetallic compounds exist can be observed. Based on the above mentioned substitutions, it can be stated that Ti and Mn are substituting Fe (located at site 1*a*) in the TiFe structure throughout the solubility homogeneity domain.[18,24]

The present study aims at investigating the effects of Mn and Ti substitution for Fe on the hydrogenation properties of TiFe alloy. To investigate and improve the hydrogenation properties of Ti(Fe,Mn)-based compounds, the percentages of alloying elements (Ti, Fe, Mn) were varied. The Mn content is varied between 0 and 5 at.%, while the Ti:Fe atomic ratio is mapped from 1:1 to 1:0.9. The results from this study, especially as concerns thermodynamic data, will be compared to previous literature data to describe in depth the hydrogen storage properties of TiFe in the selected composition range. A full picture of the dependence of hydrogen storage properties as a function of the composition will be discussed, together with some observed correlations.

## 2. Experimental

### 2.1. Sample preparation

Nine Ti-Fe-Mn samples (from s1 to s9) were synthetized with compositions given in **Table 1** and displayed in the phase diagram reported in **Figure 1**. Three Mn contents 0 (s1-s2), 2.5 (from s3 to s7) and 5 at.% Mn (s8-s9) were selected, while the Ti-content was swept, at the most, between 48.8 (s3) and 54.1 at.% Ti (s7). All the alloys were prepared by induction melting of bulk pure elements under argon in a water-cooled copper crucible. The amount of material was kept constant among the different samples at 8-10 g. The pure elements were cleaned from superficial oxides and weighted to the desired compositional ratio. Initially, a primary alloy containing only Ti (Alfa Aesar, 99.99%) and Fe (Alfa Aesar, 99.97%) had been melted. Afterwards, Mn (Goodfellow, 99.98%) was added, and the ingots were turned over and melted three more times to enhance homogeneity. Furthermore, to attain thermodynamic equilibrium, the samples were annealed in a resistive furnace at 1000 °C for 1 week and quenched to room temperature into water. To prevent contamination, the samples were previously wrapped in a tantalum foil, introduced into a silica tube and sealed under argon atmosphere. Weight losses lower than 0.1% were detected during melting and annealing. The ingots were roughly crushed under air and then finely smashed into fine powder under inert atmosphere down to submillimetre size. Argon-filled glove boxes, with a circulation purifier, were used to store, prepare and manipulate the material. $O_2$ and $H_2O$ levels were lower than 1 ppm to minimize oxidation and degradation of the material.

**2.2. Characterizations**

*2.2.1. Electron Probe Micro-Analysis*

Metallographic examination and elemental analysis by electron probe micro-analysis (EPMA, Cameca SX100) were performed to check the homogeneity and phase composition of the alloys. Precipitation of minor phases in the TiFe matrix was observed through Back-Scattered Electron (BSE) images. Acceleration voltage and beam current were set to 15 kV and 40 mA, respectively. The composition of the different phases was determined by EMPA from 10 to 100 point measurements on polished samples. The standard deviation of these measurements is given as the uncertainty of the phase composition. This uncertainty results from counting statistics, experimental setup, size of precipitates and local chemical fluctuations.

*2.2.2. Powder X-ray diffraction*

Powder X-Ray Diffraction (XRD) patterns were obtained at room temperature on a Bruker D8 Advance Bragg Brentano diffractometer, using Cu-Kα radiation (λ=1.5418 Å), in a 2θ-range of 20–120°, step size 0.02° and time step 7 seconds. All patterns were refined by the Rietveld method [34] using the FullProf package.[35]

**Table 1** – Sample list, names, nominal composition and elemental analysis (in at.%) of phases as determined by EPMA for samples s1-9.

| Sample | | Nominal Composition (at.%) | Phase | Ti (at.%) | Fe (at.%) | Mn (at.%) |
|---|---|---|---|---|---|---|
| s1 | **TiFe** | $Ti_{50.0}Fe_{50.0}$ | TiFe | 50.9±0.2 | 49.1±0.2 | 0 |
| s2 | **TiFe$_{0.90}$** | $Ti_{52.6}Fe_{47.4}$ | TiFe | 51.8±0.1 | 48.2±0.1 | 0 |
| | | | β-Ti | 80.0±0.1 | 20.0±0.1 | 0 |
| s3 | **TiFeMn$_{0.05}$** | $Ti_{48.8}Fe_{48.8}Mn_{2.4}$ | TiFe | 50.7±0.2 | 47.2±0.6 | 2.2±0.5 |
| | | | $TiFe_2$ | 36.0±0.2 | 58.7±0.4 | 5.3±0.4 |
| s4 | **TiFe$_{0.95}$Mn$_{0.05}$** | $Ti_{50.0}Fe_{47.5}Mn_{2.5}$ | TiFe | 50.8±0.1 | 46.9±0.6 | 2.3±0.5 |
| s5 | **TiFe$_{0.90}$Mn$_{0.05}$** | $Ti_{51.3}Fe_{46.2}Mn_{2.6}$ | TiFe | 51.3±0.4 | 46.5±0.5 | 2.3±0.5 |
| | | | $Ti_4Fe_2O$ | 66.7±0.2 | 31.2±0.2 | 2.1±0.1 |
| s6 | **TiFe$_{0.85}$Mn$_{0.05}$** | $Ti_{52.6}Fe_{44.7}Mn_{2.6}$ | TiFe | 51.9±0.1 | 45.6±0.3 | 2.5±0.2 |
| | | | $Ti_4Fe_2O$ | 66.8±0.1 | 31.3±0.1 | 1.9±0.1 |
| | | | β-Ti | 78.7±0.3 | 19.0±0.3 | 2.3±0.1 |
| s7 | **TiFe$_{0.80}$Mn$_{0.05}$** | $Ti_{54.1}Fe_{43.2}Mn_{2.7}$ | TiFe | 52.0±0.2 | 45.6±0.5 | 2.4±0.3 |
| | | | $Ti_4Fe_2O$ | 66.9±0.1 | 31.1±0.1 | 2.1±0.1 |
| | | | β-Ti | 78.3±0.2 | 19.2±0.2 | 2.5±0.1 |
| s8 | **TiFe$_{0.90}$Mn$_{0.10}$** | $Ti_{50.0}Fe_{45.0}Mn_{5.0}$ | TiFe | 51.0±0.2 | 43.4±0.9 | 4.6±0.1 |
| s9 | **TiFe$_{0.80}$Mn$_{0.10}$** | $Ti_{52.6}Fe_{42.1}Mn_{5.3}$ | TiFe | 51.7±0.1 | 43.9±0.5 | 4.5±0.4 |
| | | | β-Ti | 78.0±0.1 | 17.0±0.1 | 5.0±0.1 |

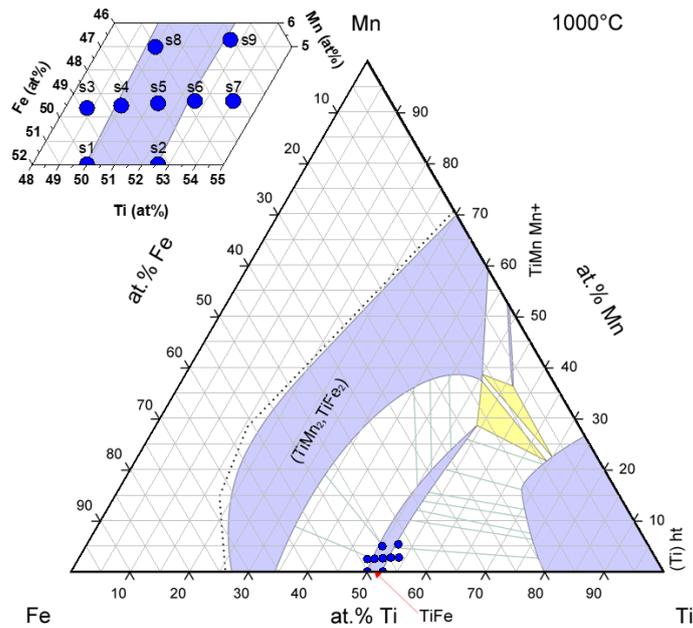

**Figure 1** – Isotherm section at 1000 °C of the Ti-Fe-Mn phase diagram and investigated compositions (dots in blue, zoomed and labelled in the up left corner).[36]

### 2.2.3. Hydrogenation properties

Activation, kinetic studies and Pressure Composition Isotherm (PCI) curves were monitored with a custom-made Sieverts' type apparatus. Approximately 500 mg of sample were loaded into a stainless steel sample holder inside the glove box, under argon atmosphere, then connected to the rig, evacuated from Ar atmosphere under primary vacuum at room temperature and activated by exposing them to pure gaseous hydrogen (Linde, 99.9999%) under different pressure and temperature conditions, depending on the investigated alloy.

Full activation of 50 at.% Ti samples (s1, s4 and s8) was quite difficult to achieve. Therefore, for these samples, a static pressure of 5.0 MPa was loaded at 25 °C. The temperature was raised uniformly within 2 hours from about 25 °C to a final value of 400 °C, held constant for 4 hours and then decreased to room temperature. Milder conditions of pressure and temperature were not successful in the activation of these samples.

PCI curves were measured after five hydrogen absorption-desorption (2.5 MPa/vacuum) cycles at 25 °C, to ensure a full activation, reproducible PCI isotherms and determine absorption rate. An oil or water thermostatic bath was used to maintain the selected temperature during PCI recording. PCI curves at temperatures comprised between 25 and 85 °C and hydrogen pressures comprised between 0.01 and 10 MPa were measured by the manometric Sieverts' method. Prior to each PCI, the sample was dehydrogenated at 200°C under primary vacuum for 30 minutes.

## 3. Results

### *3.1. Chemical, structural and morphological characterization*

**Table 1** lists the results of the EPMA analysis, showing the compositions of the different phases in the investigated alloys. For all alloys, the Ti-content of the main (matrix) phase ranges between 50.7 at.% (s3) and 52.0 at.% (s7), while Fe and Mn contents are close to the alloy nominal compositions. Therefore, this matrix is assigned to the TiFe phase. The compositional balance between Fe and Mn contents suggests that Mn substitutes Fe (site 1*a*) in the TiFe structure in agreement with previous reports.[24] In addition, thanks to BSE images (**Figure 2**), secondary phases could be detected and analysed by EPMA. The amounts of secondary phases are higher in samples s3 and s7, the overall composition of which is outside the TiFe homogeneity range (**Table 1**). The stoichiometry of secondary phases depends on Ti-content. On the Ti-poor side (sample s3), the secondary phase has a ratio of Ti to Fe (plus Mn) of about 1:2. It is therefore assigned to the intermetallic compound $TiFe_2$. On the Ti-rich side (sample s7), the secondary phase has a high Ti-content (78 at.%), which corresponds to the formation of β-Ti solid solution as corroborated by XRD (**Figure 3** and **Table 2**) analyses. In fact, the composition of this phase, which contains also Mn when present, corresponds to the high temperature allotropic form of Ti, in which large amounts of Fe can be solubilized, *i.e.* up to 20 at.%.[36] This phase could be also detected by BSE, EPMA and XRD for Ti-rich samples s2, s6 and s9. Besides, in all Ti-rich samples (s2, s5, s6, s7 and s9; >50 at.% Ti), the XRD patterns (**Figure 3**) evidence the formation of an oxide, which corresponds to a Ti to (Fe,Mn) ratio of 2:1 in the EPMA analyses. This phase can be assigned to the formation of $Ti_4Fe_2O$. In some alloys (s2, s9), it was not possible to perform reliable EPMA analysis due to reduced precipitate sizes as compared to EPMA resolution (~ 1 μm). The compositional analysis evidenced that Mn is present in $Ti_4Fe_2O$ as well.

Generally, samples on the Ti-poor side (50 at.% Ti: s1, s4 and s8) are single phase Ti(Fe,Mn) with no significant formation of oxide. Likely, precipitates are observed in BSE analysis (**Figure 2**), however their amount cannot be quantified by EPMA and XRD analysis, since precipitate size is below EPMA spatial and XRD detection limit. In contrast, for Ti-content above 50 at.%, oxide formation is detected by XRD in the range 1-4 wt.% (**Table 2**). In these samples (s2, s6, s7, s9), the β-Ti solid solution is formed as dendrites or precipitates at the grain boundaries, in the amount of 2-5 wt.%. As a general trend, Mn seems to reduce the formation of the oxide and of the β-Ti solid solution, while Ti content above 50 at.% favours the formation of the oxide, most likely due to the high reactivity of Ti towards the oxygen introduced during the alloying process or already present in the starting raw materials. The study evidenced that the presence of oxygen reactive metals (*e.g.* Ti) and possible oxygen contamination during synthesis cause the formation of $Ti_4Fe_2O$ phase, in good accordance with the Ti-Fe-O phase diagram.[36]

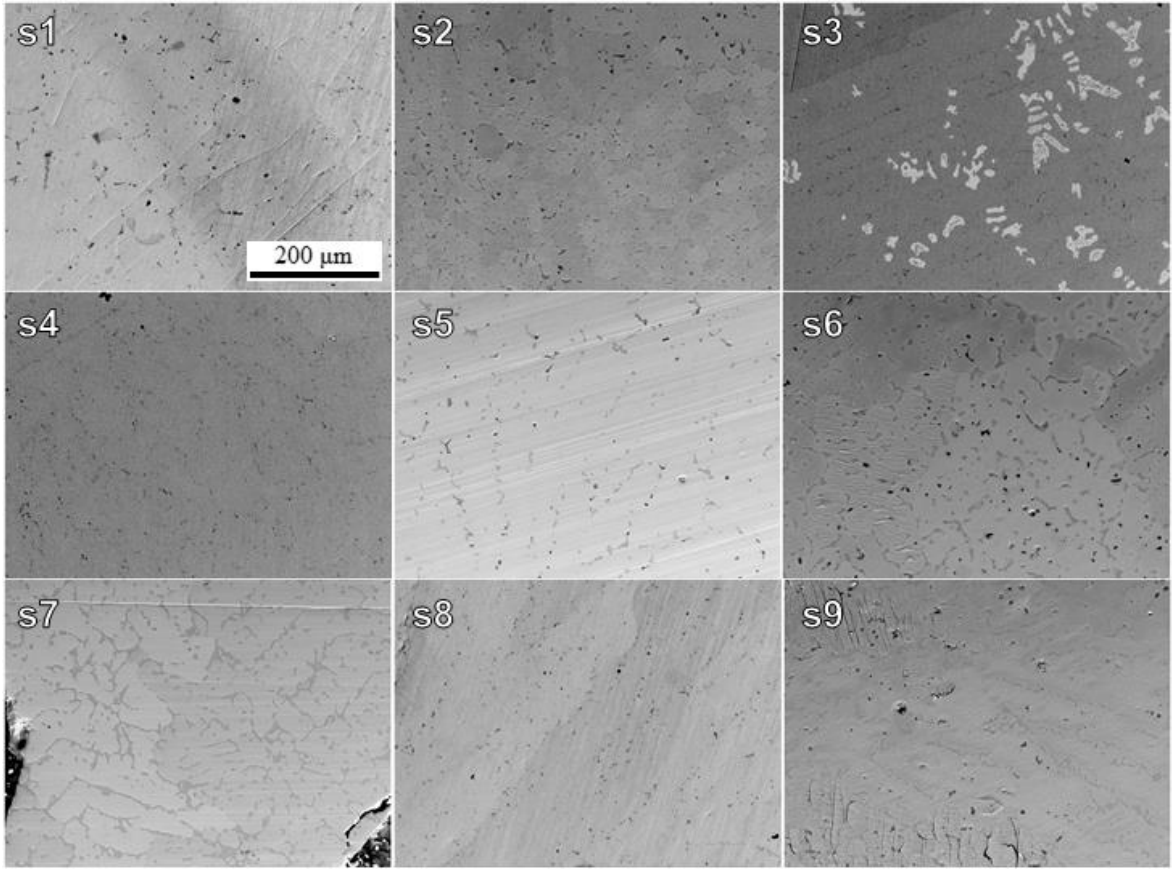

**Figure 2** – BSE metallographic images of samples s1-9.

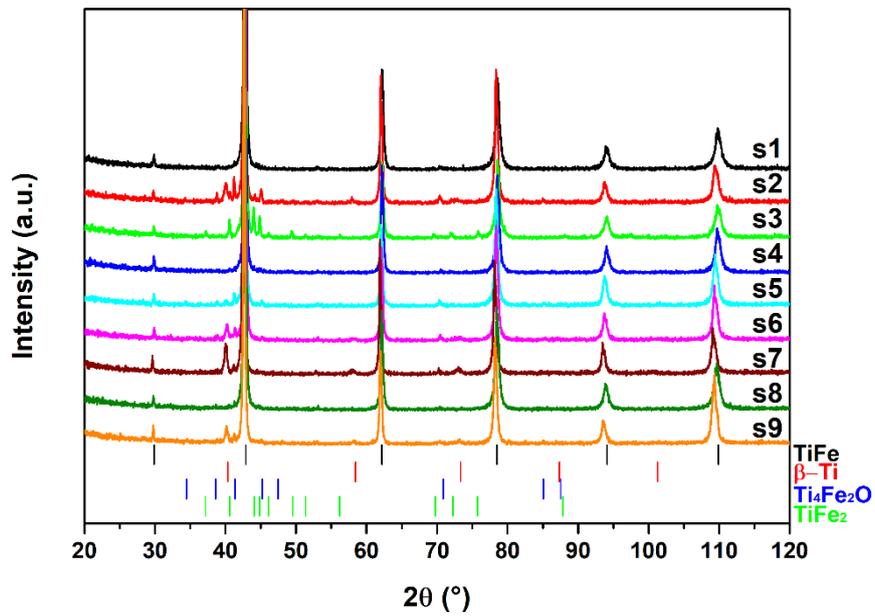

**Figure 3** – X-Ray diffraction patterns of synthetized alloys (s1-9) after annealing at 1000 °C for 1 week. Detailed data for the phases identified in the patterns are provided in **Table 1** and **Table 2**.

**Table 2** – Phase distribution and lattice parameter '*a*' of the CsCl-type TiFe phase for the s1-9 alloys synthetized in this work, as determined by Rietveld refinement of XRD data (**Figure S2**).

| Sample | | TiFe $a$ (Å) | TiFe wt.% | β-Ti wt.% | $Ti_4Fe_2O$ wt.% | $TiFe_2$ wt.% |
|---|---|---|---|---|---|---|
| **s1** | **TiFe** | 2.976(4) | 100 | | | |
| **s2** | **TiFe$_{0.90}$** | 2.982(7) | 91.0±0.8 | 5.0±0.5 | 4.0±0.5 | |
| **s3** | **TiFeMn$_{0.05}$** | 2.976(3) | 91.0±0.6 | | | 9.0±0.5 |
| **s4** | **TiFe$_{0.95}$Mn$_{0.05}$** | 2.977(0) | 100 | | | |
| **s5** | **TiFe$_{0.90}$Mn$_{0.05}$** | 2.982(9) | 98.2±0.5 | | 1.8±0.5 | |
| **s6** | **TiFe$_{0.85}$Mn$_{0.05}$** | 2.985(6) | 94.8±0.6 | 2.8±0.5 | 2.4±0.5 | |
| **s7** | **TiFe$_{0.80}$Mn$_{0.05}$** | 2.986(3) | 93.1±0.7 | 5.8±0.5 | 1.1±0.5 | |
| **s8** | **TiFe$_{0.90}$Mn$_{0.10}$** | 2.979(7) | 100 | | | |
| **s9** | **TiFe$_{0.80}$Mn$_{0.10}$** | 2.987(0) | 95.5±0.8 | 2.8±0.5 | 1.7±0.5 | |

The lattice parameters of the cubic CsCl-type TiFe phase are very close for all samples (**Table 2**) and in good agreement with literature values.[18,20,37–39] However, a careful inspection shows that a gradual increase of the lattice constant appears, following the amount of Mn and Ti. Thus, for 50 at.% Ti content, the cell parameter increases with Mn content from 2.976 Å (s1) up to 2.980 Å (s8); analogously, the cell parameter increases with Ti content from 2.977 Å (s4) up to 2.986 Å (s6) for 2.5 at.% Mn. Accordingly, the largest cell parameter is found for sample s9 (2.986 Å) with maximum Mn and Ti contents. This concurs with the fact that both, Mn and Ti, could substitute Fe in the TiFe structure and that the atomic radii of the former elements ($r_{Mn}$ = 1.35 Å, $r_{Ti}$ = 1.46 Å) are larger than that of Fe ($r_{Fe}$ = 1.24 Å). It can also be noted that the cubic cell parameter *a* of the alloys lying outside the TiFe homogeneity domain is identical to those of the alloys sitting at the corresponding phase border, *i.e.* $a_{s3} = a_{s4}$ and $a_{s6} = a_{s7}$ (**Figure 1** and **Table 2**).

*3.2. Hydrogenation properties*

*3.2.1. Activation behaviour and absorption rate*

The activation of samples with 1:1 Ti:Fe stoichiometry (Ti-poor side of the homogeneity domain; s1, s3, s4 and s8) was achieved in harsh conditions (**Table 3**). As reported before, 5 MPa of hydrogen and a heat treatment with a temperature ramp up to 400 °C were necessary for their full activation.[10,11] In contrast, almost all the Ti-rich alloys (s2, s6 and s7) could be activated at much milder conditions: room temperature (RT) and 2.5 MPa of $H_2$. Some Ti-rich alloys (s5 and s9) required moderate heating (100 °C) under higher hydrogen pressure (5 MPa). After activation, the hydrogen uptake displays fast kinetics. The time ($t_{90}$) to complete 90% of full reaction at 25°C under 2.5 MPa, was lower than 3 minutes (**Table 3** and **Figure S1**). Increasing Ti and Mn contents, the reaction kinetics tends to slightly slow down as evidenced by increasing $t_{90}$ values.

**Table 3** – Data for the activation ($T$, $P_{H_2}$, incubation time) of samples s1-9, their absorption rate ($t_{90}$) during the fifth hydrogenation cycle prior to PCI measurement at 25 °C under applied pressure of $P_{H_2}$ = 2.5 MPa (**Figure S1**), hysteresis for the first plateau at 25 °C ($ln\frac{P_{abs}}{P_{des}}$), capacity of the α phase in absorption at 25 °C, total capacity measured at 25 °C, and reversible capacity at 25 °C in the range 0.03-2.5 MPa.

| Sample | T (°C) | $P_{H_2}$ (MPa) | Inc. t (h) | $t_{90}$ (s) | $ln\frac{P_{abs}}{P_{des}}$ 1st plateau | Cap. α phase (wt.%) | Tot. cap. (wt.%) at 25 °C [P. (MPa)] | Rev. Cap. (wt.%) at 25 °C [0.03-2.5 MPa] |
|---|---|---|---|---|---|---|---|---|
| s1 | 400 | 5.0 |  |  | 0.89 | 0.04 | 1.69 [5.8] | 1.31 |
| s2 | 25 | 2.5 | 10 |  | 0.78 | 0.02 | 1.66 [5.8] | 1.54 |
| s3 | 400 | 5.0 |  |  | 0.89 | 0.04 | 1.55 [5.8] | 1.24 |
| s4 | 400 | 5.0 |  |  | 0.84 | 0.04 | 1.73 [8.0] | 1.30 |
| s5 | 100 | 5.0 | 11½ | 44 | 0.70 | 0.05 | 1.84 [5.5] | 1.72 |
| s6 | 25 | 2.5 | 7 | 106 | 0.70 | 0.08 | 1.73 [2.4] | 1.63 |
| s7 | 25 | 2.5 | 6 | 90 | 0.78 | 0.07 | 1.55 [5.7] | 1.45 |
| s8 | 400 | 5.0 |  |  | 0.78 | 0.05 | 1.75 [7.9] | 1.50 |
| s9 | 100 | 5.0 | 6¾ | 155 | 0.61 | 0.06 | 1.77 [5.5] | 1.70 |

*3.2.2. Thermodynamics*

**Figure 4** displays the PCI curves for all investigated samples (s1-9) at different temperatures. PCI curves generally consist of a first (low) and second (high) pressure plateau, both in absorption and desorption, though in some cases the double-plateau feature is not so evident due to slope feature (s5 and s9) or unachievable pressure required for observing the second plateau (s1, s3, s4 and s8). For thermodynamic analysis and comparison purposes, the values of the first and second plateau pressures were collected at 0.6 and 1.4 wt.%, which typically corresponds to half-plateau concentrations. These values were used to build the Van't Hoff plots (**Figure S3**) from which the thermodynamic properties, enthalpy Δ$H$ and entropy Δ$S$ of reaction, were determined and gathered in **Table 4**.

Clear absorption/desorption hysteresis effects are observed for all alloys. The hysteresis between absorption and desorption isotherms at room temperature is similar for all samples (**Table 3**), but it can be noticed that the Ti-poor samples (s1, s4 and s8) have larger hysteresis as compared to Ti-rich ones (s2, s6 and s9).

A clearer comparison of absorption and desorption PCI curves at 25 °C can be observed in **Figure 5**. Sample s1, TiFe, at 5.8 MPa, reaches a maximum $H_2$ capacity of 1.69 wt.% at 25 °C. Equiatomic TiFe was investigated by Reilly et al.[4] back in 1974. They reported a capacity of 1.72 wt.% at 30 °C under $P_{H2}$ = 6.3 MPa in good agreement with this study. The typical formation of the α phase (TiFeH$_{0.1}$), β monohydride (TiFeH$_{1.04}$), and γ dihydride (TiFeH$_{1.90}$) crystal structures during the loading process is in agreement with the literature, as revealed by the two plateaus. The plateau pressures concur with previous findings, in fact, the pressure of the first desorption plateau of sample s1 at 55 °C is equal to 1.045 MPa, slightly lower than the value reported by Reilly et al.[4]: 1.1 MPa. This slight difference may result from the larger cell parameter of s1 (2.976 Å) as compared to literature data (2.972 Å), as it will be discussed later on.[5]

Sample s2 (TiFe$_{0.90}$) has a rather good hydrogen capacity, equal to 1.66 wt.%, under 5.8 MPa at 25 °C. Ti-rich intermetallic compound TiFe$_{0.90}$ was recently investigated by Guéguen et al.[20]. The PCI curve at 25 °C reported in ref.[20] is well reproduced in this study.

Sample s3 (TiFeMn$_{0.05}$) has one of the lowest capacities, together with s7 (TiFe$_{0.80}$Mn$_{0.05}$), both equal to 1.55 wt.% at 5.8 MPa and 25 °C.

Back in 1981, Mintz et al.[37] investigated the composition TiFe$_{0.95}$Mn$_{0.05}$ (s4). In the present study sample s4 (TiFe$_{0.95}$Mn$_{0.05}$) reaches at 25 °C a maximum capacity of 1.73 wt.% at 8 MPa and the defined thermodynamic of the material during absorption and desorption of the first plateau well agree with Mintz et al.[37]. The second plateau was only visible at 25 °C, thus, it was impossible to define its thermodynamics.

Among all the investigated samples s5 (TiFe$_{0.90}$Mn$_{0.05}$) has the highest capacity, 1.84 wt.%, at 25 °C within the measured pressure window (0.01 to 5.5 MPa).

Sample s6 (TiFe$_{0.85}$Mn$_{0.05}$) presents a quite flat PCI curve with a maximum capacity of 1.73 wt.% at 2.4 MPa and 25 °C. This result is in good agreement with the study of Challet *et al.*[18].

Sample s8 (TiFe$_{0.90}$Mn$_{0.10}$) maximum hydrogen capacity reaches 1.75 wt.% at 7.9 MPa and 25 °C. TiFe$_{0.90}$Mn$_{0.10}$ was extensively studied in the literature and the PCI curves measured in the present investigation agree well with literature ones.[3,32,37,38,40–45] As an example, the curve reported by Leng *et al.*[43] at 4.0 MPa reaches a capacity of 1.72 wt.%, close to the value measured here.

Finally, the maximum capacity of sample s9 (TiFe$_{0.80}$Mn$_{0.10}$) is rather high, equal to 1.77 wt.% at 5.5 MPa and 25 °C. It locates in between the value reported in literature by Guéguen *et al.*[20], *i.e.* 1.68 wt.% (25 °C, 2.5 MPa), and the other one reported by Challet *et al.*[18], *i.e.* 1.92 wt.% (25 °C, 2.0 MPa). The absorption pressure of the PCI at 25 °C is slightly higher than in earlier works. For example, at 1.4 wt.%, an equilibrium pressure of 0.338 MPa is recorded, which is slightly higher than 0.18 MPa in ref.[20].

The thermodynamics of the system of samples s1-9 are summarized in **Table 4**, while capacity of the α phase and the total capacity of the material at maximum pressure as well as the reversible capacity investigated at 25 °C are gathered in **Table 3**.

Thermodynamic values related to the second plateau of samples s1 and s4 could not be determined because the plateau pressure values are too high to be properly measured (**Figure 4**).

Absorption enthalpy values of the first plateau range from -28.3 to -24.3 kJ mol$^{-1}$, while the entropy change ranges between -100 and -95 J mol$^{-1}$ K$^{-1}$. The second plateau enthalpies and entropies of absorption present higher values compared to the absorption first plateau ones, ranging between -35.4<$\Delta H$<-26.2 kJ mol$^{-1}$ and -132<$\Delta S$<-114 J mol$^{-1}$ K$^{-1}$. Upon desorption this trend is maintained. However, higher values are evidenced compared to the absorption ones: 27.4<$\Delta H$<32.4 kJ mol$^{-1}$ and 96<$\Delta S$<107 J mol$^{-1}$ K$^{-1}$ for the first plateau and 35.2<$\Delta H$<38.9 kJ mol$^{-1}$ and 124<$\Delta S$<144 J mol$^{-1}$ K$^{-1}$ for the second one.

In the literature, the enthalpy and entropy absolute values of absorption and desorption reactions for TiFe (s1) are dispersed, ranging from 28 to 31 kJ mol$^{-1}$ ($\Delta H$) and 52 to 118 J mol$^{-1}$ K$^{-1}$ ($\Delta S$) for the first plateau, and ranging from 32 to 34 kJ mol$^{-1}$ ($\Delta H$) and 125 to 137 J mol$^{-1}$ K$^{-1}$ ($\Delta S$) for the second plateau.[4,11,16,46]. Furthermore, previous thermodynamic studies evidenced a wide range of enthalpy and entropy values for s8 too, *i.e.* in absorption -24.2<$\Delta H$<-21.8 kJ mol$^{-1}$ and -96<$\Delta S$<-89 J mol$^{-1}$ K$^{-1}$, and 27.6<$\Delta H$<30.3 kJ mol$^{-1}$ and 100<$\Delta S$<110 J mol$^{-1}$ K$^{-1}$ in desorption.[37,38,40,42] Values determined in this work are in agreement with previous findings, apart from the high values recorded for s8 second plateau during desorption: $\Delta H$=38.9 kJ mol$^{-1}$ and $\Delta S$=144 J mol$^{-1}$ K$^{-1}$ (**Table 4**).

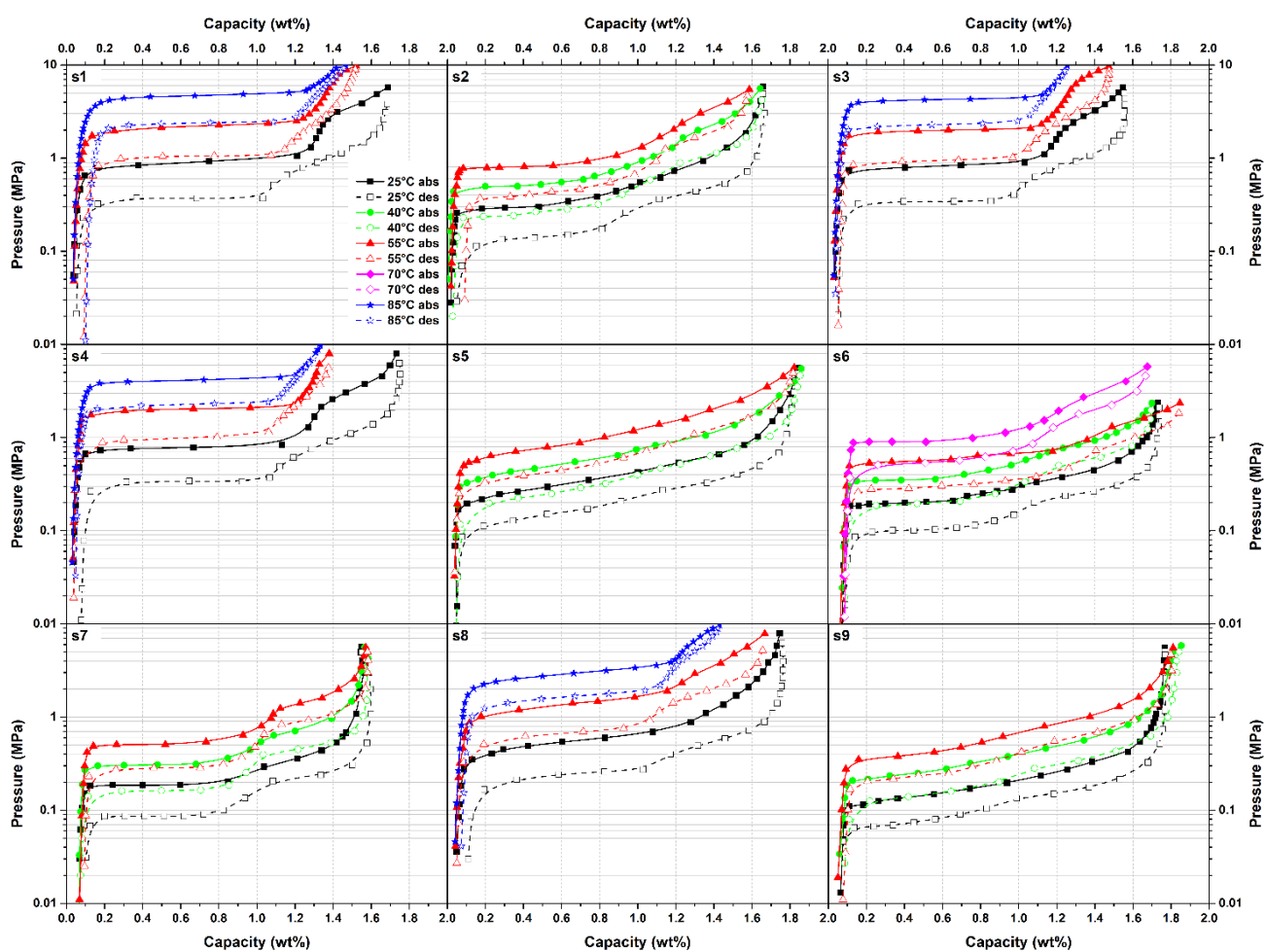

**Figure 4** – Absorption (full symbols) and desorption (empty symbols) PCI curves for samples s1-9; at 25 °C (black squares), 40 °C (green circles), 55 °C (red triangles), 70 °C (purple diamonds) and 85 °C (blue stars). Solid lines in absorption and dashed lines in desorption are a guide for the reader. The error bars are within the data points.

**Table 4** – Hydrogenation thermodynamics of samples s1-9 determined from Van't Hoff plots (**Figure S3**) of PCI curves at different temperatures.

Plateau pressure data were collected for hydrogen contents of 0.6 (first plateau) and 1.4 wt.% $H_2$ (second plateau).

| Sample | Plateau | Absorption | | | | | | | Desorption | | | | | | |
|---|---|---|---|---|---|---|---|---|---|---|---|---|---|---|---|
| | | Plateau Pressure | | | | | Enthalpy | Entropy | Plateau Pressure | | | | | Enthalpy | Entropy |
| | | MPa | MPa | MPa | MPa | MPa | kJ mol$^{-1}$ | J mol$^{-1}$ K$^{-1}$ | MPa | MPa | MPa | MPa | MPa | kJ mol$^{-1}$ | J mol$^{-1}$ K$^{-1}$ |
| T | | 25°C | 40°C | 55°C | 70°C | 85°C | | | 25°C | 40°C | 55°C | 70°C | 85°C | | |
| s1 | first | 0.894 | | 2.169 | | 4.620 | -24.3 | -100 | 0.368 | | 1.045 | | 2.345 | 27.4 | 103 |
| s2 | first | 0.329 | 0.555 | 0.842 | | | -25.5 | -95 | 0.151 | 0.281 | 0.439 | | | 29.0 | 101 |
| s3 | first | 0.833 | | 2.022 | 4.307 | | -24.3 | -99 | 0.343 | | 0.949 | 2.303 | | 28.2 | 105 |
| s4 | first | 0.786 | | 2.010 | 4.171 | | -24.7 | -100 | 0.341 | | 0.994 | 2.290 | | 28.2 | 105 |
| s5 | first | 0.317 | 0.521 | 0.833 | | | -26.2 | -97 | 0.157 | 0.245 | 0.446 | | | 28.2 | 98 |
| s6 | first | 0.212 | 0.376 | 0.585 | 0.936 | | -27.8 | -99 | 0.105 | 0.206 | 0.305 | 0.556 | | 30.6 | 103 |
| s7 | first | 0.187 | 0.316 | 0.518 | | | -27.6 | -98 | 0.086 | 0.164 | 0.284 | | | 32.4 | 107 |
| s8 | first | 0.535 | | 1.367 | | 2.838 | -24.7 | -97 | 0.245 | | 0.676 | | 1.641 | 28.1 | 102 |
| s9 | first | 0.152 | 0.276 | 0.431 | | | -28.3 | -98 | 0.083 | 0.156 | 0.243 | | | 29.2 | 96 |
| s1 | second | 2.930 | | | | | | | 1.060 | | | | | | |
| s2 | second | 1.073 | 2.285 | 3.468 | | | -31.9 | -127 | 0.491 | 1.130 | 1.954 | | | 37.5 | 139 |
| s3 | second | 3.194 | | 8.382 | | | -26.2 | -116 | 1.126 | | 4.146 | | | 35.3 | 139 |
| s4 | second | 2.551 | | | | | | | 0.947 | | | | | | |
| s5 | second | 0.656 | 1.165 | 2.067 | | | -31.1 | -120 | 0.342 | 0.656 | 1.259 | | | 35.3 | 128 |
| s6 | second | 0.460 | 0.930 | 1.060 | 3.008 | | -32.5 | -121 | 0.266 | 0.597 | 0.680 | 2.034 | | 35.2 | 126 |
| s7 | second | 0.503 | 1.043 | 1.851 | | | -35.4 | -132 | 0.262 | 0.558 | 1.099 | | | 38.9 | 138 |
| s8 | second | 1.220 | | 3.552 | | 9.082 | -29.7 | -120 | 0.543 | | 2.055 | | 7.567 | 38.9 | 144 |
| s9 | second | 0.338 | 0.615 | 1.064 | | | -31.1 | -114 | 0.185 | 0.375 | 0.683 | | | 35.4 | 124 |

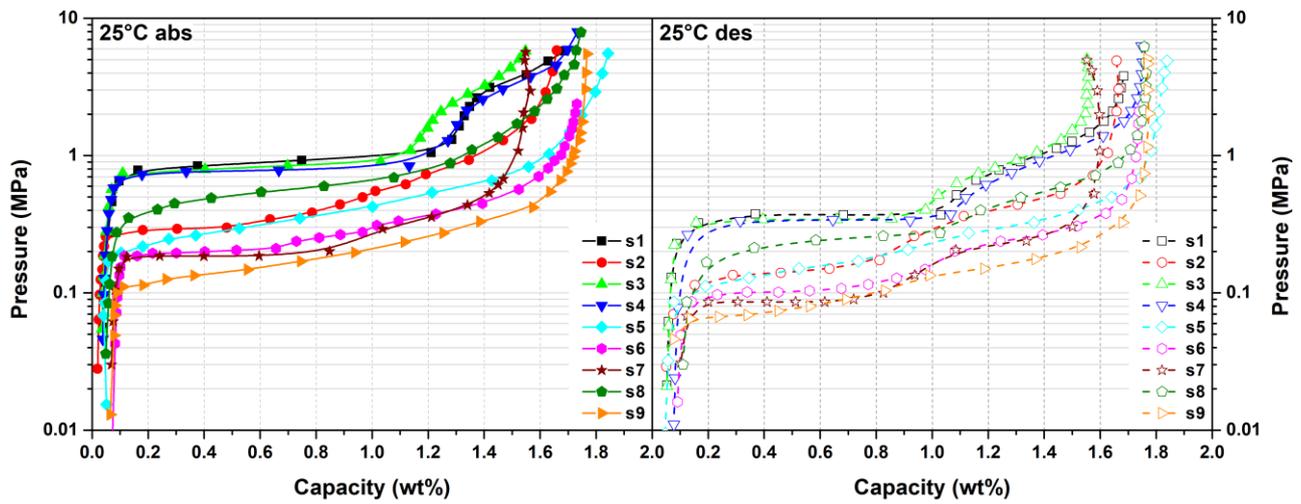

**Figure 5** – Comparison of absorption (left) and desorption (right) PCI curves for samples s1-9 at 25 °C. The error bars are within the data points.

## 4. Discussion

In the following section, discussion of the presented results will be based on the main properties and how they can be tuned and modified by tailoring the composition. The discussion will consider the composition of the main phase (equiatomic TiFe or not, addition of Mn) and the role of secondary phases.

### *4.1. Hydrogenation properties*

### *4.1.1. Activation behaviour and absorption rate*

The presence of small amounts of secondary phases, such as β-Ti solid solution and $Ti_4Fe_2O$, plays a role in the activation process, easing it.[47] They facilitate the penetration and the first hydrogen absorption after some hours of incubation time (**Table 3**). The amount of oxide secondary phase reactive to hydrogen could be a crucial factor for an easy activation process; the effortful activation of samples s5 and s9 could be related to an insufficient content of such secondary phases. Indeed, Ti-poor alloys, which are mostly single phase (**Figure 2** and **Table 2**), required very harsh condition of activation.

Precipitates that can readily be hydrogenated could act as fast diffusion path for hydrogen, helping the activation process to occur at lower temperature and pressure conditions.[25,47] This study evidences that an amount of secondary phases close to 5 wt.% can be a good trade off in terms of easy activation and satisfactory H$_2$ capacity of the material (samples s2, s6 and s7).

An exception to this general trend is sample s5. It contains a low amount of oxide phase but rather easy activation. The non-equiatomic composition of the main phase could be another key parameter that guarantees easier activation process, suggesting a key role of Ti and Mn substitution on the Fe site in enhancing the activation. Ti higher affinity for H can potentially justify this behaviour. On the other hand, absorption rates are affected by substitution, which causes a slightly higher value for t$_{90}$. Nevertheless, absorption rates are fast and kept under 3 minutes.

### 4.1.2. Thermodynamics

#### 4.1.2.1. Capacity

Compounds with high maximum capacities (s4, s5, s6) contain minor amount of secondary phases. Although s1 has the highest TiFe content (100% by XRD, minor precipitates observed by BSE), and a lower maximum capacity of 1.69 wt.%, this is most likely due to the fact that the maximum gas pressure available was not sufficient to reach the full capacity.

The overall compositions of samples s2 and s7 are outside the TiFe homogeneity range (**Figure 1**). Therefore, significant amounts of secondary phases are formed: TiFe$_2$ and mainly β-Ti for s3 and s7 alloys, respectively. TiFe$_2$ does not react with hydrogen in the experimental conditions, whereas β-Ti reacts irreversibly, reducing the reversible capacity of the material.[4,23] Consequently, the presence of β-Ti and TiFe$_2$ leads to lower capacities.

The lower hydrogen capacity of sample s6, compared to s5, could be justified by the higher Fe concentration of β-Ti, since the amount of the intermetallic compound TiFe with 94.8 wt.% is significantly lower than the 98.2 wt.% of s5. The lower capacity of sample s7 relates to the high amount of β-Ti as well.

The high capacity of s5 concurs with low amount of secondary phases. Only the oxide $Ti_4Fe_2O$ was detected below 2 wt.%. This phase has been reported to react reversibly with hydrogen to a certain extent.[25,47] In fact, for sample s2, despite the low TiFe concentration of 91 wt.%, a capacity of 1.66 $H_2$ wt.% is still reached, which is higher than the capacity of sample s7 (1.55 $H_2$ wt.%) that has only 1.1 wt.% $Ti_4Fe_2O$.

As introduced, it can be observed that the amount of hydrogen stored in the α phase is higher for samples s6, s7, and s9 (**Table 3**), reducing the reversible capacity. On the other hand, low capacities can be possibly explained by a too high equilibrium pressure, but mainly by the presence of inactive or irreversible hydrogenated phases, as in the case of s3 and s7 (**Table 3**). High capacities are usually observed in samples that contain both Mn and Ti substitution for Fe, such as s5, s8, and s9. These samples also have the highest amount of TiFe phase (**Table 2**) and rather flat PCI curves with low equilibrium pressure, if compared to sample s1 (TiFe) and s4 ($TiFe_{0.95}Mn_{0.05}$) that, even if single phase, have the two plateaus at rather high equilibrium pressures.

In conclusion, with the intent of Ti(Fe,Mn) alloy applications for stationary hydrogen storage, **Table 3** presents the calculated reversible capacity between 0.03 MPa (in desorption) and 2.5 MPa (in absorption) at 25 °C. Reported reversible capacities are higher in Ti and Mn substituted-alloys, thanks to their lower PCI equilibrium pressure and flatter plateau. Samples on the Ti-rich side combine easier activation as well.

*4.1.2.2.* **Hysteresis**

The lowest hysteresis factor is obtained for sample s9, which has the highest amount of both Ti and Mn. The hysteresis of s7 is significantly smaller than the one of s1, this could be related with the difference in cell parameter, *i.e.* 2.986 Å for s7, much higher than the one of s1 (2.977 Å).

Interestingly, the hysteresis decreases with Ti and Mn contents, lowering with the enlargement of the TiFe cell parameter (**Figure 6**). Higher content of Ti and Mn promotes a narrower hysteresis and a higher TiFe cell parameter, when going from s1 to s9 the cell volume changes significantly. As it can be observed in **Figure 7**, owing to the larger radius of Ti with respect to Mn, it is evident that Ti substitution (s1 to s2; s8 to s9; and from s3 to s7) leads to higher volume expansion if compared to the effect of Mn in enlarging the cell volume (s1, to s4 and s8; or s2 to s6 and s9).

This trend, evidenced in **Figure 6**, seems valid for most samples except for s7, which could be related to some influence from secondary β-Ti precipitates. Inverse dependence between hysteresis and TiFe cell volume could be related to relative strains induced by expansion during the alloy-to-hydride transition. In addition, this is related to the mechanical properties of the material, which however should be verified by nucleation study (for instance by in-situ Transmission Electron Microscopy) to understand how such strain depend on lattice parameters within the two-phase field.

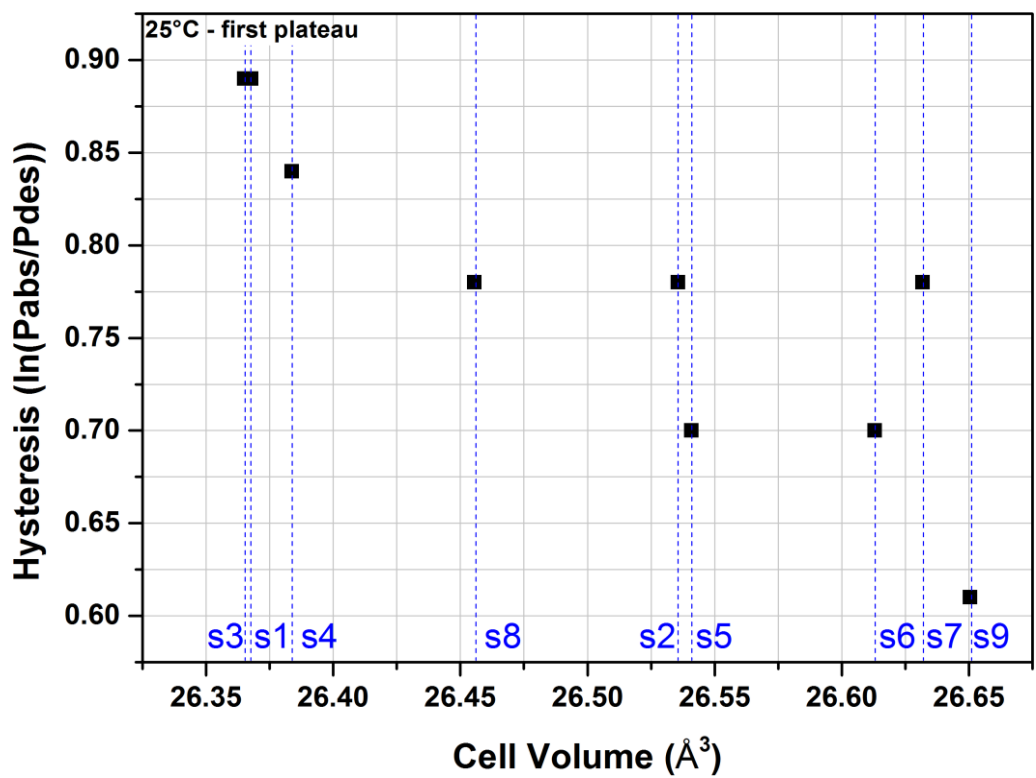

**Figure 6** – Hysteresis as a function of TiFe-phase cell volume for the investigated alloys (s1-9). The error bars are within the data points.

*4.1.2.3.* **Plateau pressure and slope**

Sample s2 (TiFe$_{0.90}$) has lower plateau pressures compared to sample s1 (TiFe), thus evidencing the effective role of Ti substitution for Fe in enlarging the cell parameter and lowering the plateau.

In comparison to the equiatomic TiFe alloy (s1), a slight reduction of the plateau pressure is detected in s4, owing to Mn substitution. Sample s8 (TiFe$_{0.90}$Mn$_{0.10}$) has the highest quantity of Mn among the 1:1 stoichiometry samples. Its equilibrium pressure is the lowest among equiatomic samples (s1, s4 and s8), owing to the higher amount of Mn and thus higher cell parameter of TiFe phase.

Focusing on the elemental substitution in TiFe-alloys, this work evidences a rather powerful influence of the Ti substitution for Fe in Ti-rich samples that has the effect of lowering equilibrium pressure. This effect is distinctly recognizable from looking at the PCI curves from samples s3 to s7, which all contain about 2.5 at.% Mn; the same applies for the PCI curves of s1 and s8 compared to s2 and s9 respectively. Instead, the impact of Mn substitution on PCI curves is modest; always tend towards lower values of pressure when a higher amount of Mn is introduced. This effect is visible comparing sample s1 to s4 and s8, and comparing s2 to s6 and s9.

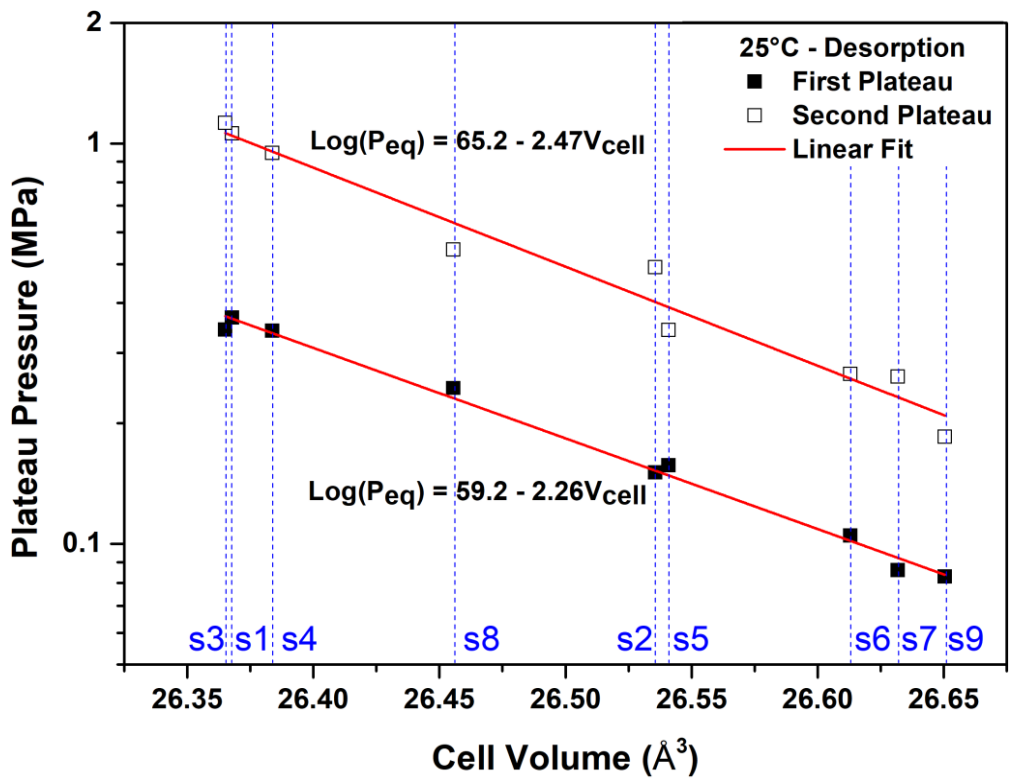

**Figure 7** – Linear correlation between first and second plateau pressure in desorption at 25 °C and the TiFe-phase cell volume of investigated alloys (s1-9). The error bars are within the data points.

**Figure 7** shows that a linear correlation can be observed between equilibrium pressure (in logarithmic scale) and cell volume of the TiFe phase. As an early conclusion, this study evidences the greater influence of Ti in lowering pressure rather than Mn, clearly related to geometrical effects, owing to the larger radius of Ti ($r_{Ti}$(1.46 Å) > $r_{Mn}$(1.35 Å)) that allows extensive volume expansion by the addition of small concentration. When moving from s1 to s2, thus increasing the Ti amount in the TiFe phase from 50.9 to 51.8 at.%, the cell parameter variates from 2.976 to 2.982 Å. For practical applications, the linear correlation allows to estimate which cell volume of the alloy will be needed to obtain a given equilibrium pressure in desorption in TiFe-based system, which can be furthermore related to Mn and Ti content. Both Mn and Ti substitutions fulfil the geometric model.[48]

Considering the whole picture of the alloys investigated in this study (**Figure 5**), it can be observed that samples s1, s3, and s4 present alike PCI curve behaviours with two distinct plateaus. Generally, the two plateaus are better resolved during desorption, for all samples (**Figure 4** and **Figure 5**).

Two plateaus are clearly visible in s2, s4 and s7 PCI curves, as for s1. On the other hand, sample s5 (TiFe$_{0.90}$Mn$_{0.05}$) displays a single sloped plateau. Indeed, the addition of Mn leads to a smoothing effect of the plateaus for absorption and desorption, meaning that two separate plateaus can no longer be clearly identified. This is in good agreement with the literature.[5,33]

Another example of sloped plateau during hydrogenation is sample s8. The PCI curve has a marked slope and consequently the two plateaus associated to the distinct formation of the β and γ phase are not clearly identified, contrary to binary TiFe. Interestingly, the marked slope in PCI curve of sample s8 can be correlated with chemical fluctuations, which are evidenced by the high standard deviation of Fe content determined by EPMA (**Table 1**).

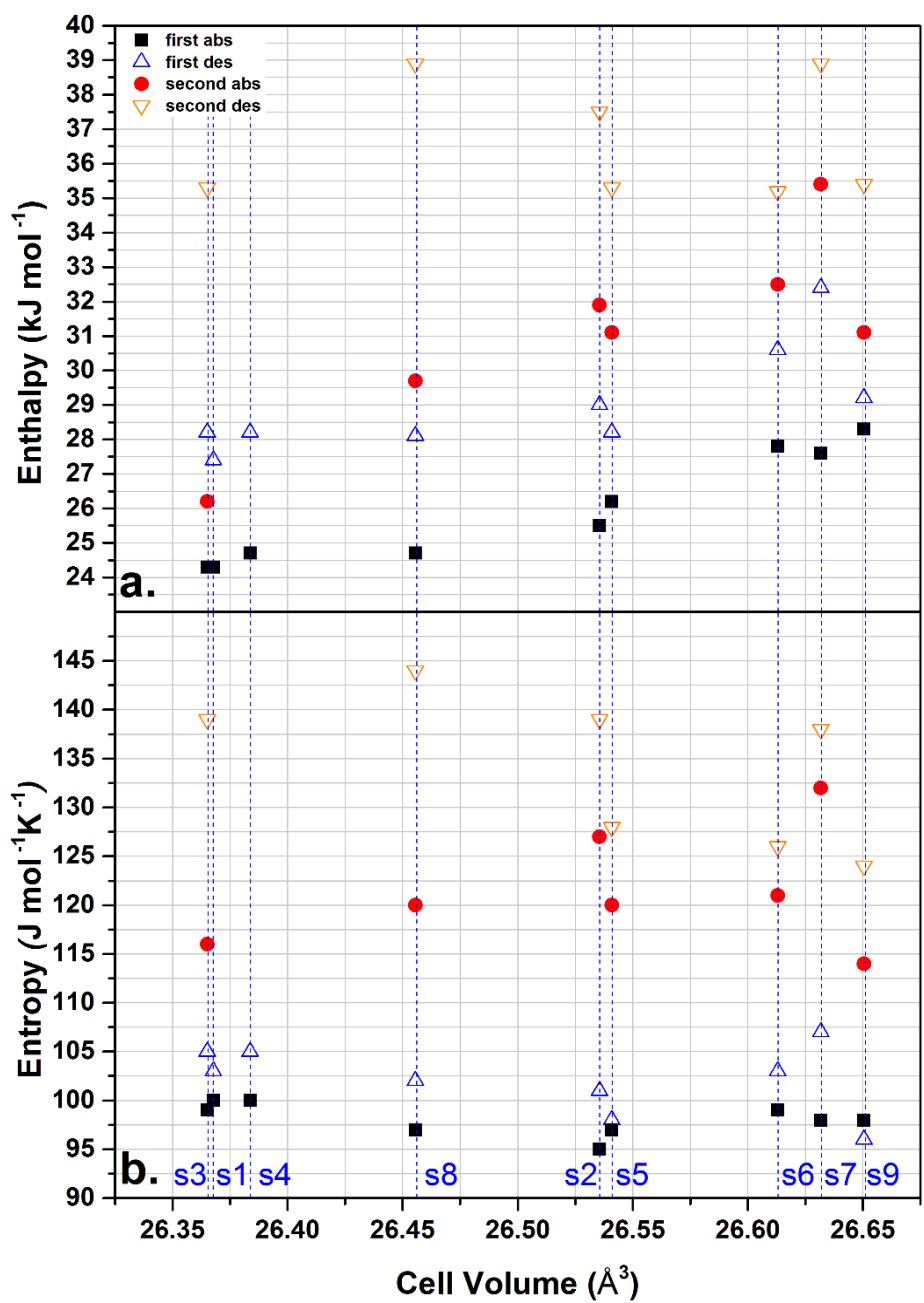

**Figure 8** – Absolute thermodynamic values determined by the Van't Hoff plot for first and second plateau in absorption and desorption of tne investigated alloys (s1-9). Enthalpy (a) and entropy (b) as a function of TiFe-phase cell volume. The error bars are within the data points.

### 4.1.2.4. Hydrogenation enthalpy and entropy

This work presents a detailed investigation and determination of enthalpies and entropies of reaction related to absorption and desorption of both, first and second plateau, in a wide range of composition, *i.e.* variating the Ti and Mn contents (**Table 4**). The determined thermodynamics generally agree well with previous literature. Slight differences are found between absorption and desorption thermodynamics, while sensible changes in thermodynamic values are recorded between the first and second plateau. The absolute enthalpy of the first plateau (24-32 kJ mol$^{-1}$) is lower than that of the second one (26-39 kJ mol$^{-1}$), being slightly lower for absorption than for desorption. The absolute entropy values, both for absorption and desorption, are lower for the first (95-107 J mol$^{-1}$ K$^{-1}$) than for the second (114-144 J mol$^{-1}$ K$^{-1}$) pressure plateau as well.

Generally, Ti-rich samples (with enlarged cell-volume) have higher values of reaction enthalpy (hydride stabilization) both in absorption and in desorption (**Figure 8-a**). It can be evidenced that the introduction of Mn in 1:1 stoichiometry samples (s1, s4, s8) does not change significantly the thermodynamics, while on moving from s1 to s2 (Mn-free samples) and from s3 to s7 (2.5 at.% constant Mn content) the hydrogenation enthalpy of the first plateau significantly increases with Ti substitution. The enthalpy of the first hydrogenation changes dramatically when both Ti and Mn are introduced, as for s8 to s9; while in Mn-free sample it is less marked (s1 to s2).

In absorption, the hydrogenation enthalpy gradually increases with the expansion of the cell volume (Mn and Ti substitution, **Figure 8-a**). This relates to the fact that hydrides get more stable on increasing cell volume; *i.e.* fulfil the empirical geometric relationship previously proposed. This evolution seems to be verified in absorption, while it is not so clear in desorption.

The definition of a possible correlation for enthalpy values of the second plateau is less evident, due to the higher degree of slope and higher equilibrium pressure of the second plateau in some samples. Entropy values are stable over the samples (**Figure 8-b**), while they increase from first plateau in absorption to first plateau in desorption, then to second plateau absorption, finally showing the higher values in the second desorption plateau. This variation in entropy might be related to possible stress-strain effect related to hydride formation, and it has been recently reported that hydrogenation entropy linearly increases with bulk modulus of the alloy and its working function.[49]

*4.1.2.5.* **Correlations**

General trends and correlation between composition and hydrogenation as well as structural properties in all investigated samples are resumed in the counter colour maps reported in **Figure 9**. In details, substitution of Ti and Mn for Fe in TiFe-type alloy results in enlargement of the cell parameter and volume of the TiFe-phase (**Figure 9-a**). This reduces the equilibrium pressure, diminishes the pressure gap between first and second plateau and decreases the hysteresis of the isotherms (**Figure 9-c**). Thermodynamic trends of enthalpy and entropy of reactions can be appreciated in the compositional counter colour maps in **Figure 9** from **e** to **h**. It can also be seen that the samples at the Ti-rich side of the phase diagram display enhanced storage reversible capacities (**Figure 9-b**), as long as limited amount of β-Ti is formed. When high number of β-Ti precipitates is formed, it irreversibly traps hydrogen at very low pressure (**Figure 9-d**).

## 5. Conclusions and outlook

The chemical and microstructural characteristics of nine different composition in the Ti-Fe-Mn system have been systematically investigated in the present work.

The study investigates the role of Ti and Mn substitution in TiFe-type alloy in the range from 48.8 at.% to 54.1 at.% Ti, and from 0 to 5.3 at.% Mn. Within these boundaries, according to the ternary Ti-Fe-Mn phase diagram, the main phase is TiFe. Additionally, the formation of $TiFe_2$ can occur at the Ti-poor side, while at the Ti-rich side the formation of β-Ti-type and oxide phase ($Ti_4Fe_2O$-type) is observed. BSE analyses showed that the induction melting process and subsequent annealing produced alloys formed by a chemically homogeneous TiFe-type phase and minor precipitate phases. Through XRD-analyses and Rietveld's refinement, it could be determined that the resulting phases are in good agreement with the expected results based on the ternary phase diagram, and some consideration about hydrogen properties and geometrical consideration has been formulated.

The determination of the hydrogenation properties in a wide range of composition evidenced the role of microstructure and secondary phases formation on activation and kinetics. The formation of secondary phases reactive to hydrogen enhances the activation properties and allows the first hydrogenation of the material in milder conditions, as in the case of Ti and Mn substituted materials.

However, high amounts of secondary phases reduce the total and reversible capacity of the material, thus a compromise must be reached.

Thermodynamic investigations were made to examine the influence of the Ti and Mn content on hydrogenation properties. The performed PCI curves allowed a good estimation of the thermodynamic properties of the samples that agree well with the available literature, but further complete the dependency of hydrogenation properties over extended Ti and Mn compositional range. The reversible capacity determined at 25 °C between the pressure window of 0.03 and 2.5 MPa is higher in samples s5, s6, and s9, owing to reduced equilibrium pressure and a flatter plateau. In particular, sample s6 (TiFe$_{0.85}$Mn$_{0.05}$) results to be the best compositional trade-off, as it retains a reversible capacity as high as 1.63 wt.% (at 25°C and between 0.03-2.5 MPa) combined with easy activation process (6 hours incubation time under 2.5 MPa at 25 °C) and fast absorption rate, thanks to its good kinetics, with a t$_{90}$ lower than 2 minutes.

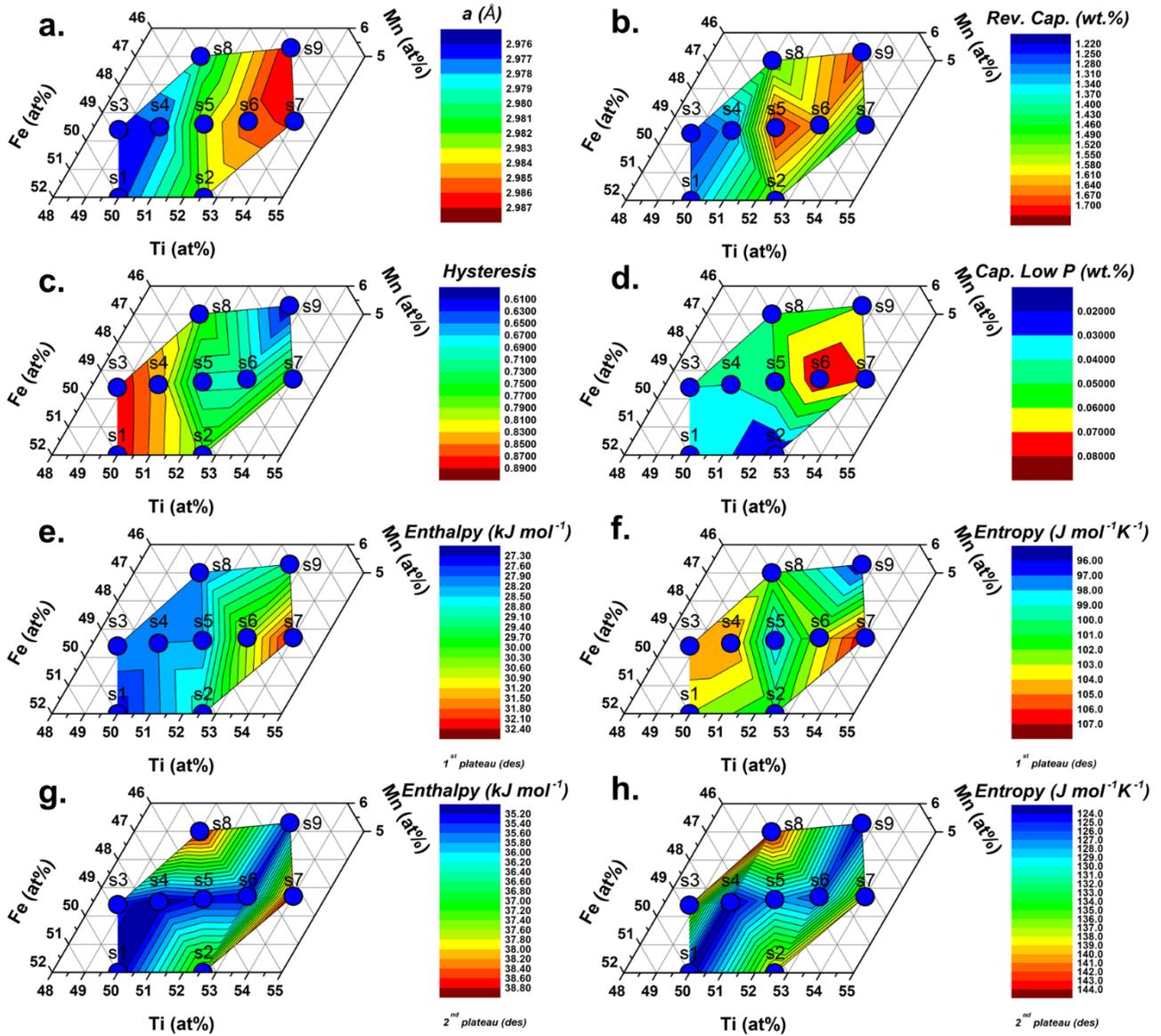

**Figure 9** – Compositional counter colour maps reporting sample s1-9 measured properties: a) cubic cell parameter (Å) of the TiFe-type phase; b) reversible capacity at 25°C between 0.03-2.5 MPa (wt.%); c) hysteresis of the first plateau; d) hydrogen stored/trapped at low pressure, i.e. below the first plateau pressure (wt.%); first plateau desorption e) enthalpy (kJ mol$^{-1}$) and f) entropy (J mol$^{-1}$ K$^{-1}$); and second plateau desorption g) enthalpy (kJ mol$^{-1}$) and h) entropy (J mol$^{-1}$ K$^{-1}$).

**Acknowledgement**

This project has received funding from the Fuel Cells and Hydrogen 2 Joint Undertaking (JU) under grant agreement No 826352, HyCARE project. The JU receives support from the European Union's

Horizon 2020 research and innovation programme and Hydrogen Europe and Hydrogen Europe Research.

The authors wish to thank E. Leroy for EPMA analysis and F. Couturas for his help with hydrogenation experiments.

This work was supported by a doctoral scholarship from the Bonn-Rhein-Sieg University of Applied Sciences (Germany)

**Electronic Supporting Information Description**

ESI includes curves of pressure as a function of time with details on activation and kinetics, Rietveld refinements, Van't Hoff plots used for the determination of thermodynamics.

**Supplementary data**

Supplementary data to this article can be found online at https://doi.org/10.5281/zenodo.4299023 and https://doi.org/10.5281/zenodo.4299000.

# ESI

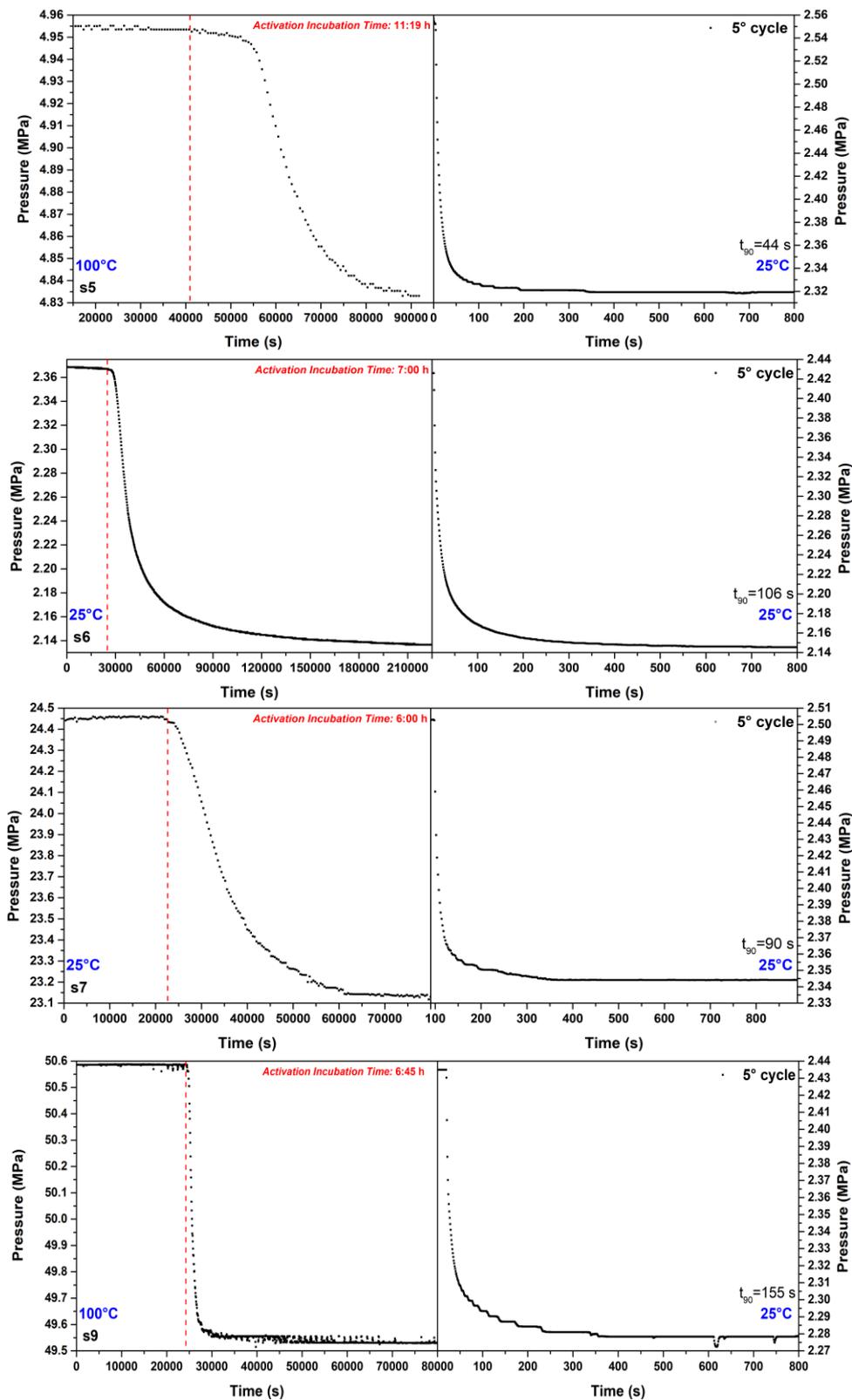

**Figure S1** – Pressure data as a function of time during first hydrogenation (activation) and during the fifth absorption after initial cycling, representative of the kinetic of the samples (s5-s6-s7-s9). Temperature, activation incubation time, and time to reach 90% of reacted fraction, $t_{90}$, are displayed as well.

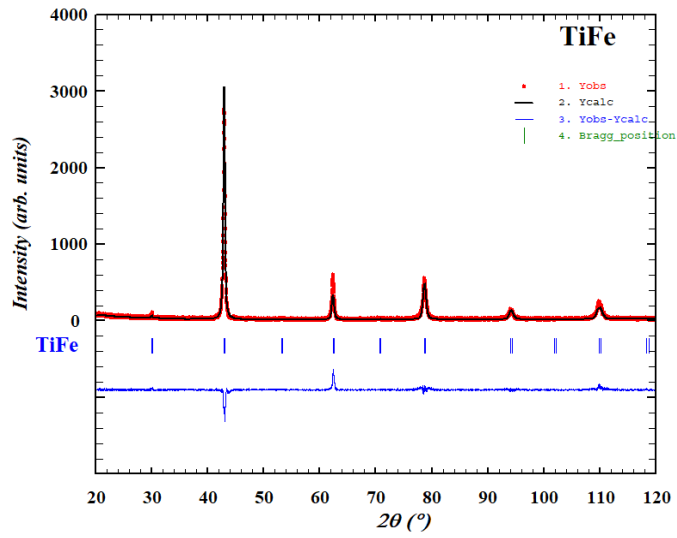
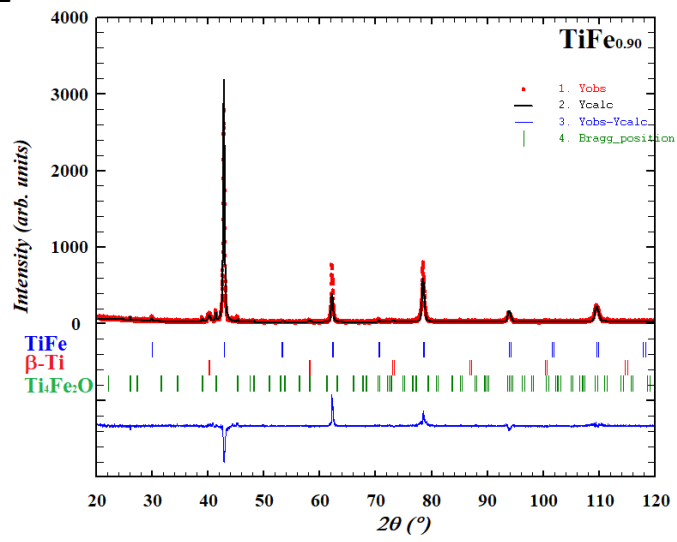
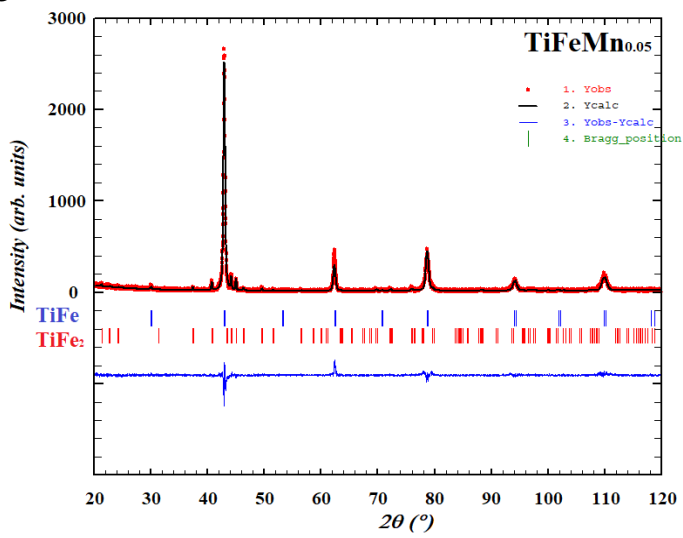

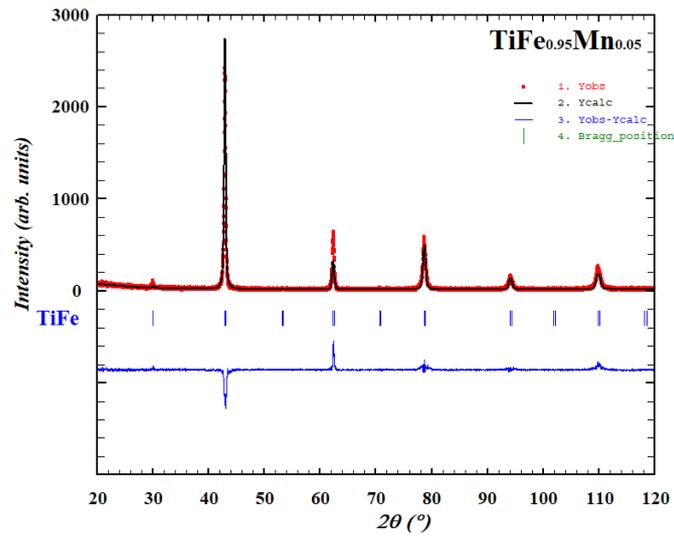

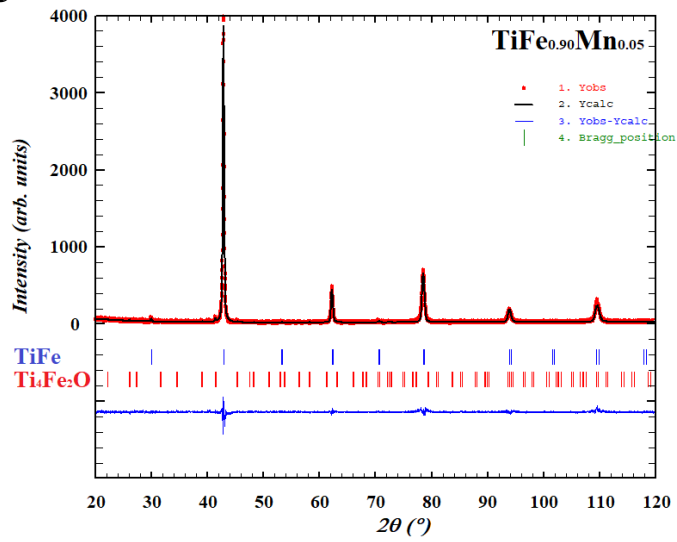

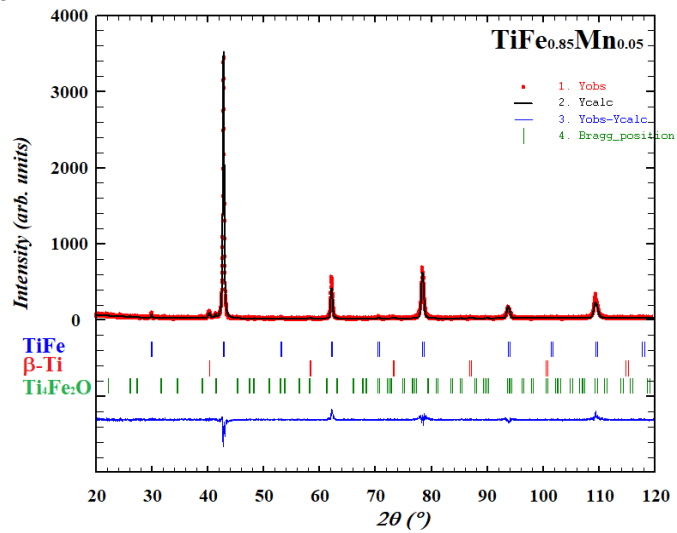

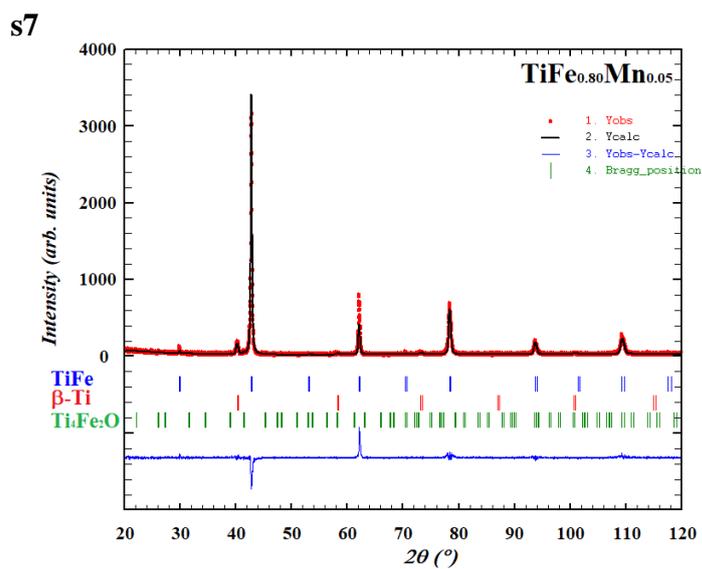
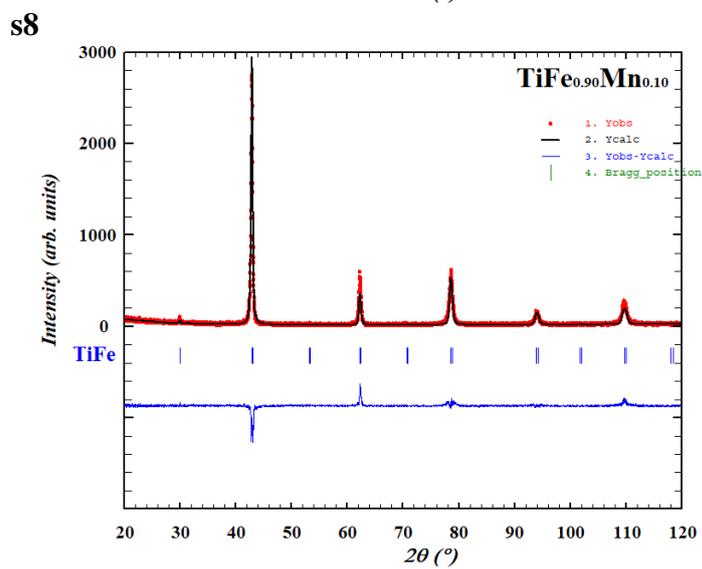
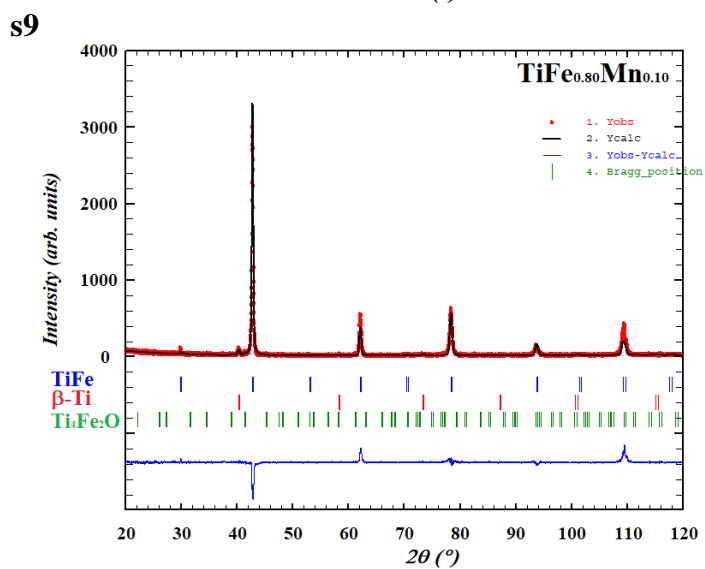

**Figure S2** – Rietveld refinements for investigated sample after annealing (s1-9).

**s1**

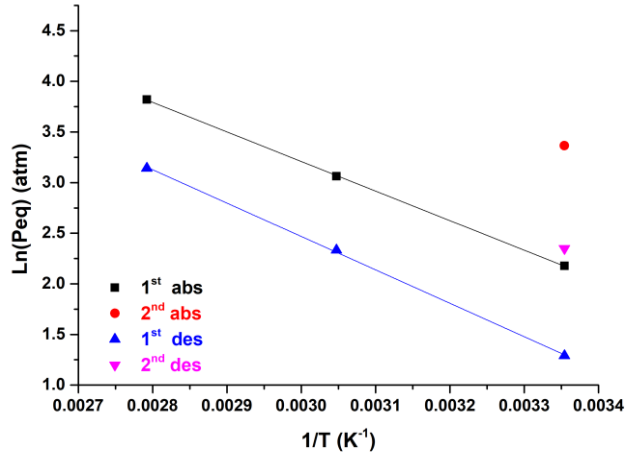

**s2**

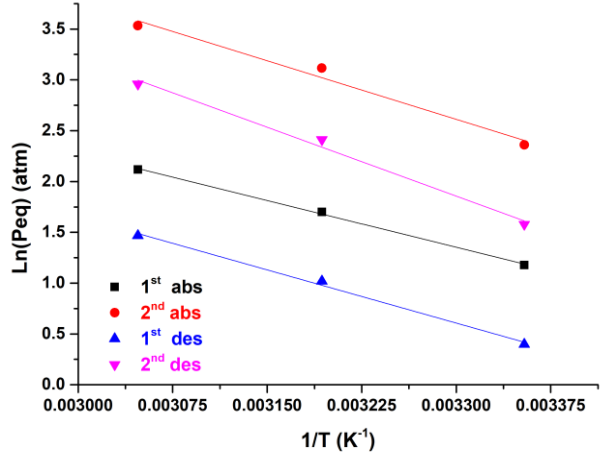

**s3**

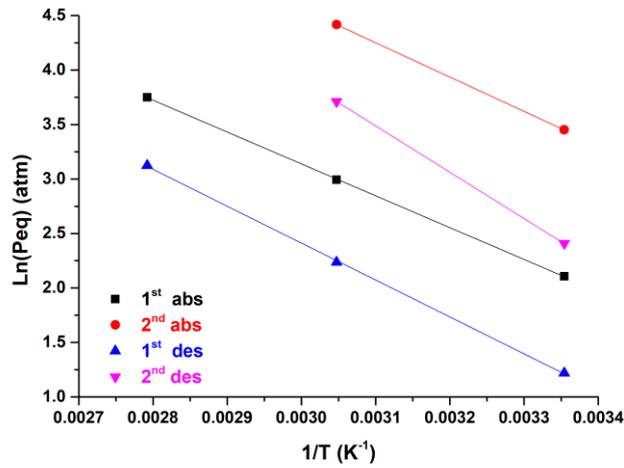

**s4**

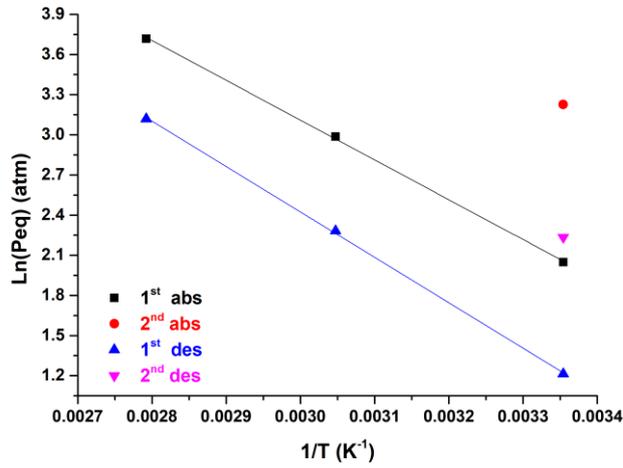

**s5**

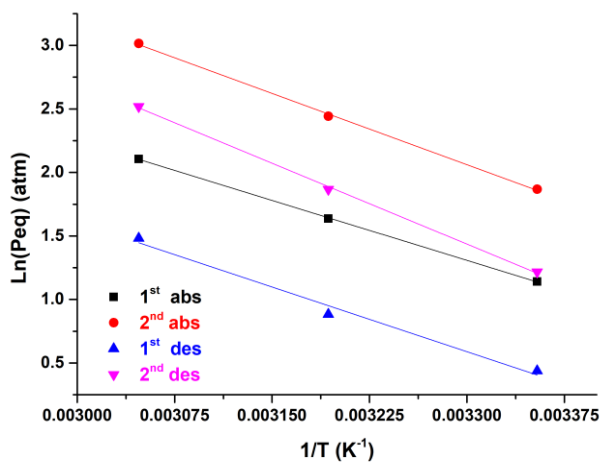

**s6**

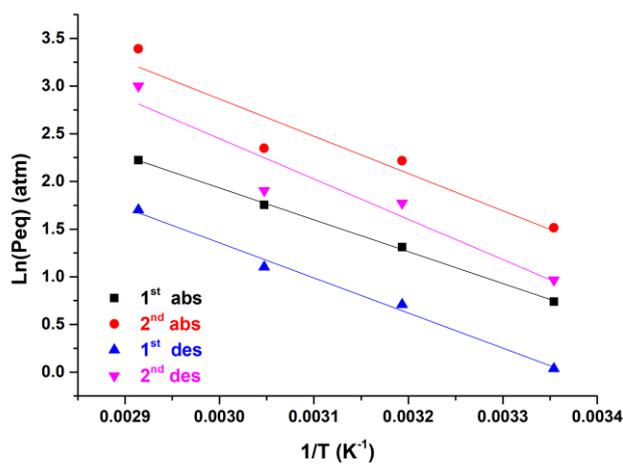
38

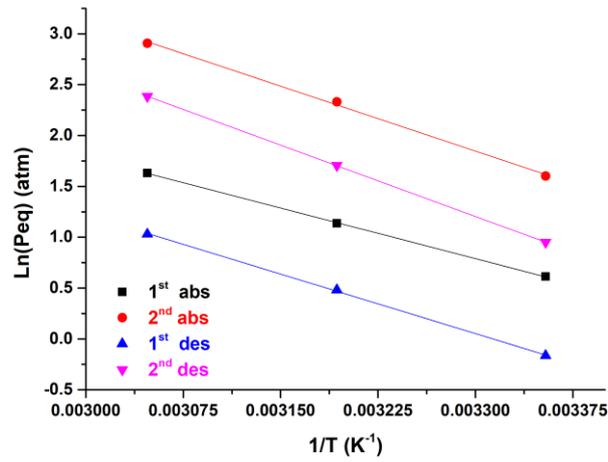

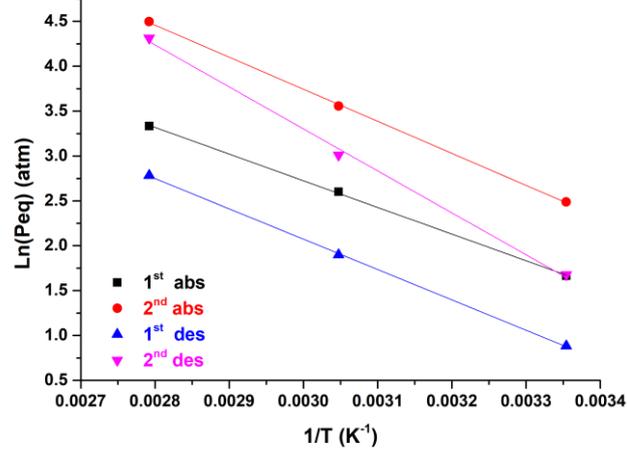

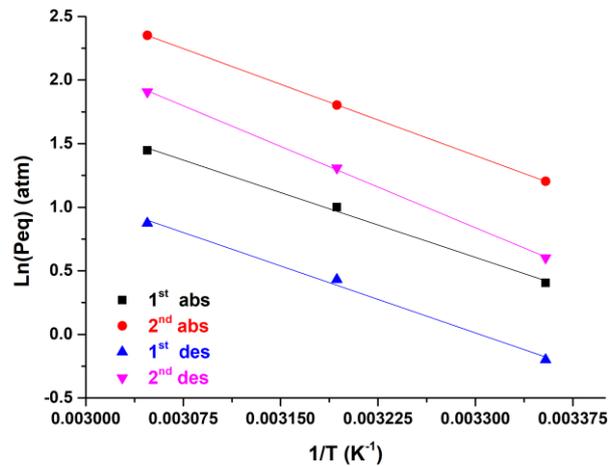

**Figure S3** – Van't Hoff plots for investigated samples s1-9.